%% file: main.tex
\documentclass[conference]{IEEEtran}
\IEEEoverridecommandlockouts

\usepackage{cite}
\usepackage{amsmath,amssymb,amsfonts}
\usepackage{algorithmic}
\usepackage{graphicx}
\usepackage{textcomp}
\usepackage{xcolor}
\usepackage{float}
\usepackage{subcaption}
\usepackage{booktabs}
\usepackage{amsmath}
\usepackage{multirow}
\usepackage{tikz}
\usepackage{hyperref}
\newcommand\circled[1]{\tikz[baseline=(char.base)]{
\node[shape=circle,draw,inner sep=1pt] (char) {#1}}}
\def\cyl#1{{#1}}

\def\BibTeX{{\rm B\kern-.05em{\sc i\kern-.025em b}\kern-.08em
    T\kern-.1667em\lower.7ex\hbox{E}\kern-.125emX}}
\begin{document}

\title{
\cyl{Efficient CPU-GPU Collaborative Inference for MoE-based LLMs on Memory-Limited Systems}
}

\author{\IEEEauthorblockN{En-Ming Huang*}
\IEEEauthorblockA{
\textit{National Taiwan University}\\
Taipei, Taiwan \\
r13922078@csie.ntu.edu.tw
}
\and
\IEEEauthorblockN{Li-Shang Lin*}
\IEEEauthorblockA{
\textit{National Tsing Hua University}\\
Hsinchu, Taiwan \\
lslin@cs.nthu.edu.tw}
\and
\IEEEauthorblockN{Chun-Yi Lee$^\dagger$}
\IEEEauthorblockA{
\textit{National Taiwan University}\\
Taipei, Taiwan \\
cylee@csie.ntu.edu.tw
}
}

\maketitle
\begin{abstract}

\cyl{Large Language Models (LLMs) have achieved impressive results across various tasks, yet their high computational demands pose deployment challenges, especially on consumer-grade hardware. Mixture of Experts (MoE) models provide an efficient solution through selective activation of parameter subsets, which reduces computation requirements. Despite this efficiency, state-of-the-art MoE models still require substantial memory beyond typical consumer GPU capacities. Traditional offloading methods that transfer model weights between CPU and GPU introduce latency, limiting inference performance. This paper presents a novel CPU-GPU collaborative inference framework that incorporates an expert caching mechanism on the GPU to reduce data transfer requirements and enable faster inference through cache hits. Computations are offloaded to CPU for efficient cache miss handling, which benefits from CPU multithreading optimizations. The evaluations of our framework demonstrate performance improvements and highlight the potential of CPU-GPU collaboration to maximize hardware utilization for single-request inference scenarios on consumer-grade systems. The implementation of our framework is available at \href{https://github.com/elsa-lab/MoE-CPU-GPU-Collaborative-Inference}{github.com/elsa-lab/MoE-CPU-GPU-Collaborative-Inference}.}

\end{abstract}

\let\thefootnote\relax\footnotetext{*Authors contributed equally. $^\dagger$ Corresponding author.\\ Accepted by the Asia and South Pacific Design Automation Conference (ASP-DAC), 2026}

\input{sections/1_introduction.tex}
\input{sections/2_motivation.tex}

\input{sections/3_method.tex}

\input{sections/4_exp.tex}
\input{sections/5_conclusion.tex}
\input{sections/6_acknowledgements}

\pagebreak
\bibliography{refs}
\bibliographystyle{IEEEtranS_our}

\end{document}

%% file: sections/1_introduction.tex
\section{introduction}
\label{sec:introduction}

\begin{figure}
    \centering
    \includegraphics[width=0.4\textwidth]{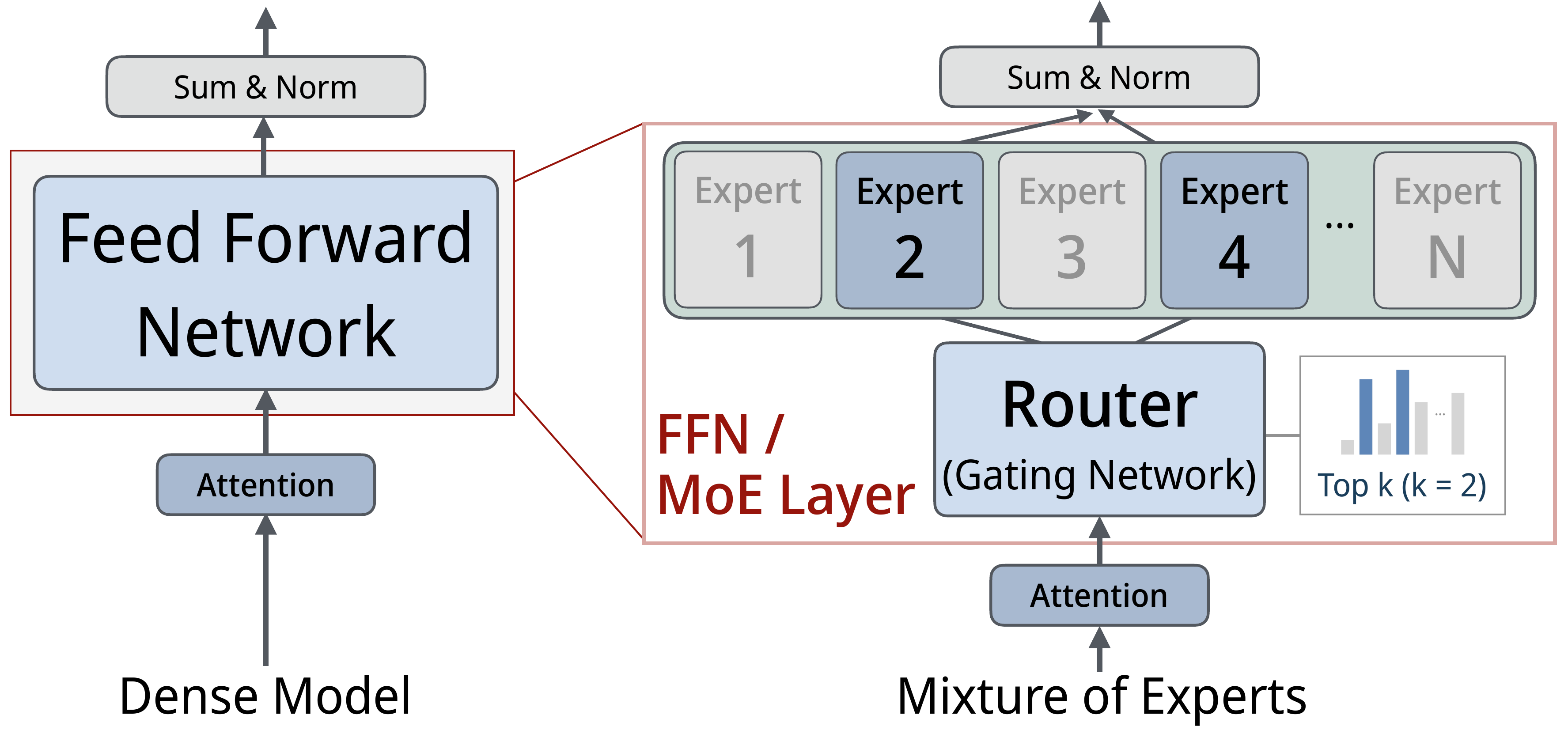}
    \caption{
    \cyl{Comparison of dense FFN and MoE layer architectures.}
    }
    \label{fig:dense-moe-architecture}
    \vspace{-1em}
\end{figure}

\cyl{Large language models (LLMs) have demonstrated remarkable effectiveness in addressing diverse real-world challenges~\cite{openai2020gpt3,openai2024gpt4,dubey2024llama3herdmodels}. Although increased model sizes have led to improved accuracy, they have also introduced substantial computational challenges. Mixture of Experts (MoE) model architectures~\cite{2017moe,fedus2022switchmoe,jiang2024mixtralexperts,2024phi3.5,mosaic2024dbrx,yang2024qwen2technicalreport,mixtral8x22b} have emerged as a computationally efficient paradigm that maintains superior performance while substantially reducing inference costs through the activation of sparse experts. For instance, Mixtral 8x7B~\cite{jiang2024mixtralexperts} achieved performance superior to that of LLaMA2-70B~\cite{2023llama2} across various benchmarks while requiring only one-fifth~\cite{jiang2024mixtralexperts} of the computational resources. In these architectures, the large feed-forward networks (FFNs) within each Transformer block~\cite{attention} are decomposed into multiple smaller expert models. As depicted in Fig.~\ref{fig:dense-moe-architecture}, the large dense FFN is replaced by an MoE layer, wherein the router network determines which experts to activate based on the input data. This selective mechanism forms a fundamental advancement in computational efficiency.}

\cyl{Despite reduced computational complexity through sparsity, MoE models encounter memory constraints on consumer-grade hardware (e.g., Mixtral 8x7B requires 80GB of model weights vs RTX 4090's 24GB capacity). To address this limitation, researchers have proposed various offloading methods that store model weights in CPU memory or SSDs and transfer weight portions to the GPU during computation under different strategies~\cite{deepspeedinference,2023moefastinf, 2024pregatedmoe,2024adapmoe,2024expertflow, xue2024moeinfinity,2023flexgen,lee2024infinigen}. On-demand expert fetching transfers specific experts based on router decisions, yet model weight transfers cannot proceed until router decisions are determined~\cite{huggingfaceTransformers, deepspeedinference}. As a result, prefetching techniques were proposed to overlap processor computations with memory transfers through expert model speculation~\cite{2023moefastinf, 2024pregatedmoe,2024adapmoe}. For instance, required experts in layer-$i+1$ are predicted in prior layers (e.g., layer-$i$), which allows earlier weight fetching. Alternatively, increased batch sizes have proven effective in enhancing total inference throughput through expert weight reuse~\cite{2024expertflow, xue2024moeinfinity,2023flexgen,lee2024infinigen}. However, minimum latency remains limited by the communication time needed for expert weight transfers.}

\begin{figure*}[t]
    \centering
    \begin{subfigure}[t]{1.0\textwidth}
        \centering
        \includegraphics[width=\textwidth,page=1]{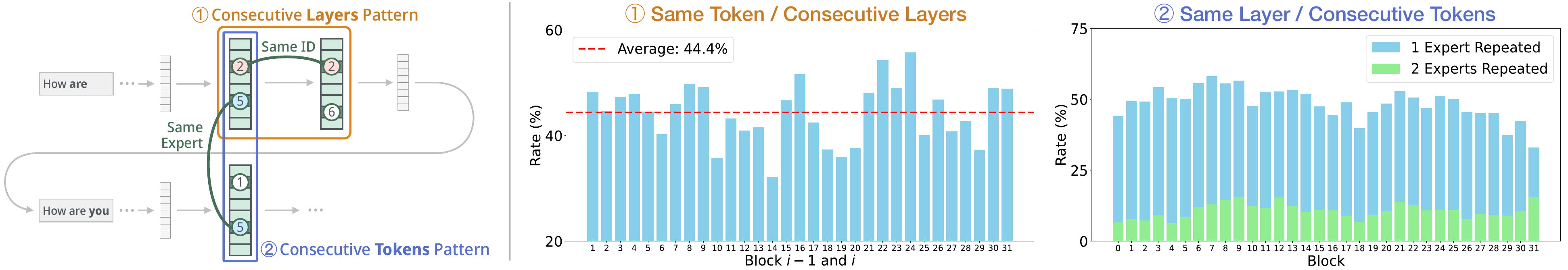}
    \end{subfigure}
    \caption{\cyl{Mixtral 8x7B (32 Transformer blocks) expert selection patterns on the MMLU~\cite{hendryckstest2021,hendrycks2021ethics} dataset: (1) Consecutive Layers Pattern and (2) Consecutive Token Pattern. Each Transformer block's router demonstrates clear patterns of expert reuse. \vspace{-1em}}
    }
    \label{fig:expert-pattern-overview}
\end{figure*}

\begin{figure}[t]
    \centering
    \includegraphics[width=0.48\textwidth]{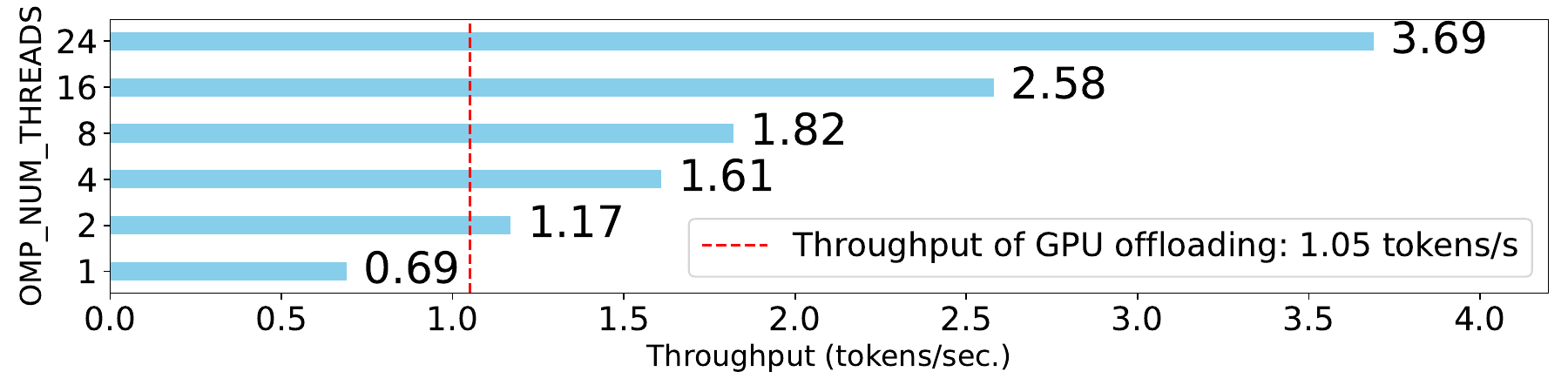}
    \vspace{-0.8em}
    \caption{\cyl{Mixtral 8x7B token generation speed vs. CPU cores.}
    }
    \label{fig:motivation-token-speed}
\end{figure}

\cyl{To address these memory and latency challenges, this paper proposes CPU-GPU collaboration for efficient inference on consumer-grade hardware, particularly for per-request scenarios common in such systems. The primary insight stems from our observation that previous research on MoE-based LLM inference has predominantly focused on GPU-based computations, which has not fully utilized CPU computational resources and therefore remains limited by model transfer time between CPU and GPU. Despite its often overlooked potential, CPU possesses significant parallel computation capabilities supported by PyTorch~\cite{pytorch}, a widely used library that implements multithreading through 
OpenMP~\cite{openmp} and vector instructions. This parallelization is controlled by the environment variable \texttt{OMP\_NUM\_THREADS}, which adjusts CPU core utilization for computation. As shown in Table~\ref{table:motivation-comparison} and Fig.~\ref{fig:motivation-token-speed}, our analysis reveals that CPU computation can achieve superior efficiency compared to GPU offloading for certain operations in the Mixtral 8x7B model. 
This advantage stems from the interplay between computation and communication costs. Although the GPU exhibits substantially faster computational speed (0.3 ms versus 25.5 ms with two threads), the significant communication overhead (28.0 ms) negates this benefit in per-request scenarios. The throughput analysis in Fig.~\ref{fig:motivation-token-speed} further suggests that CPU configurations with more than two threads offer superior throughput as compared to GPU offloading.}

\cyl{To demonstrate that expert computations can be effectively conducted on CPU rather than through GPU weight transfers, Table~\ref{table:motivation-comparison} compares transfer times and CPU computation times on an AMD Ryzen Threadripper 7960X CPU and NVIDIA GeForce RTX 4090 GPU. The communication time in the CPU version reflects output transfer from the GPU-computed attention block back to CPU. The results show that two CPU cores can surpass GPU offloading performance for the Mixtral 8x7B model. Fig.~\ref{fig:motivation-token-speed} illustrates overall token generation speed under different CPU core configurations. The results further indicate that employing at least two CPU cores achieves greater efficiency than conventional MoE layer GPU offloading. These profiled findings reveal that prefetching methods~\cite{2023moefastinf, 2024pregatedmoe,2024adapmoe} cannot achieve effective computation-communication overlap, as MoE expert transfer costs substantially exceed computation time. As a result, prefetching performance improvements remain constrained by inherent communication latency.}

\begin{table}[t]
    \centering
    \caption{
    \cyl{Communication \& computing time for Mixtral 8x7B FFN using AMD 7960X CPU and NVIDIA RTX 4090 GPU with PCIe 4.0 ×16 bandwidth (64 GB/s bidirectional).}
    }
    \label{table:motivation-comparison}
    \resizebox{0.48\textwidth}{!}{
        \begin{tabular}{l|c|cccccc}
            \toprule
            Device             & GPU   & \multicolumn{6}{c}{CPU (w/ different no. of threads)}   \\
            \midrule
            OMP\_NUM\_THREADS  & -   & 1     & 2     & 4     & 8     & 16    & 24   \\
            \midrule
            Computing time (ms.)        & 0.3  & 44.1  & 25.5  & 18.3  & 15.8  & 11.0 & 7.3 \\
            Communication time (ms.)    & 28.0 & \multicolumn{6}{c}{0.1} \\
            \bottomrule
        \end{tabular}
    }
    \vspace{-1em}
\end{table}

\cyl{Based on CPU computational capabilities and experimental insights, we introduce a CPU-GPU collaborative inference framework for MoE-based LLMs. This framework incorporates GPU expert caching and introduces novel asynchronous cache miss handling to minimize latency penalties. When required experts are not cached on GPU, computations are offloaded to CPU while experts are asynchronously fetched to GPU for subsequent token generation.
Since GPU cannot accommodate all MoE layer experts, we consider GPU memory as a cache for expert subsets. We observe expert reuse patterns across consecutive token generations in MoE models. Real-case study evidence on Mixtral 8x7B model appears in Fig.~\ref{fig:expert-pattern-overview}, which shows selected experts are re-employed in subsequent token generation with $30\%$ to $50\%$ probability. Performance comparisons in Table~\ref{table:motivation-comparison} and Fig.~\ref{fig:motivation-token-speed} further demonstrate that employing CPU to complement non-cached layer computations offers an effective solution. This methodology enables synergistic CPU-GPU resource utilization that achieves throughput surpassing GPU-only computation. Unlike prior works requiring model alteration or dataset profiling~\cite{2024pregatedmoe, 2023moefastinf,fiddler2025}, ours allows out-of-the-box operation with any MoEs.}

\cyl{We evaluate the framework performance on state-of-the-art (SoTA) MoE-based LLM models: Mixtral 8x7B~\cite{jiang2024mixtralexperts} and Phi-3.5-MoE~\cite{2024phi3.5} using an AMD Ryzen Threadripper 7960X CPU and NVIDIA GeForce RTX 4090 GPU. Our solution surpasses all previous methods~\cite{huggingfaceTransformers, deepspeedinference, 2024pregatedmoe, 2024adapmoe,fiddler2025} without accuracy trade-offs, including approaches that required model alterations~\cite{2024pregatedmoe} and dataset profiling~\cite{fiddler2025}. Specifically, our approach achieves up to $4.8$ and $10.4$ tokens per second for Mixtral 8x7B and Phi-3.5-MoE respectively in single-request scenarios, demonstrating a $4.4\times$ speedup against the prior prefetching-based approach~\cite{2024pregatedmoe}. The main contributions are:}
\begin{enumerate}
    \item \cyl{A CPU-GPU collaborative inference framework for MoE-based LLMs on resource-constrained consumer-grade systems that addresses single-request scenarios common in real-world cases with limited batch sizes. Comprehensive experiments on state-of-the-art MoE models validate our framework's effectiveness without requiring model alterations or additional prior profiling.}
    \item \cyl{The demonstration of the untapped potential of CPU computational resources through multi-threading capabilities that surpass conventional GPU-only methods. }
    \item \cyl{An innovative \textit{asynchronous expert caching mechanism} that exploits data locality in GPU memory and leverages CPU multi-core resources during cache misses while enabling asynchronous expert weight transfers to GPU.}
\end{enumerate}

%% file: sections/2_motivation.tex
\section{\cyl{Related Work and Motivational Insights}}
\label{sec:motivation}


\subsection{\cyl{Related Work}}

\cyl{To address deployment challenges of large MoE-based LLMs on memory-limited GPUs, existing offloading strategies exhibit fundamental limitations across distinct approaches. Deepspeed Inference~\cite{deepspeedinference} and Huggingface Accelerate~\cite{huggingfaceTransformers} provide on-demand expert weight fetching based on router decisions without requiring model modifications, but lack prefetching mechanisms that create significant communication latency. Throughput-oriented methods including ExpertFlow~\cite{2024expertflow} and MoE-Infinity~\cite{xue2024moeinfinity} enhance multi-user performance through token scheduling, expert prefetching, and quantization but prioritize overall system throughput over individual request latency. Edge-focused approaches such as EdgeMoE~\cite{yi2023edgemoe} and SwapMoE~\cite{2024-swapmoe} target severely constrained devices through extensive model modifications and offline profiling for quantization and pruning strategies. Router modification strategies exemplified by Pre-gated MoE~\cite{2024pregatedmoe} and AdapMoE~\cite{2024adapmoe} pre-select experts for subsequent layers based on current activations. These methods require custom prefetch layers and router logic modifications that reduce compatibility with standard inference frameworks. Router mispredictions can stall GPU execution while awaiting weight transfers, introducing substantial communication overhead. Heterogeneous system approaches such as Fiddler~\cite{fiddler2025} utilize both CPU and GPU resources through offline profiling to cache frequently activated experts based on skewed activation distributions. This profiling-dependent methodology limits generalization across MoE architectures and employs greedy algorithms with computational overhead of $O(N \times 2^N)$, creating inefficiency for larger models. Our methodology addresses these limitations by requiring no model architecture modifications, quantization, or extensive offline profiling while supporting diverse MoE-based LLM models with maintained compatibility and accuracy.}

\begin{figure*}[t]
    \centering
    \begin{subfigure}[t]{.66\textwidth}
        \centering
        \includegraphics[width=\textwidth,page=1]{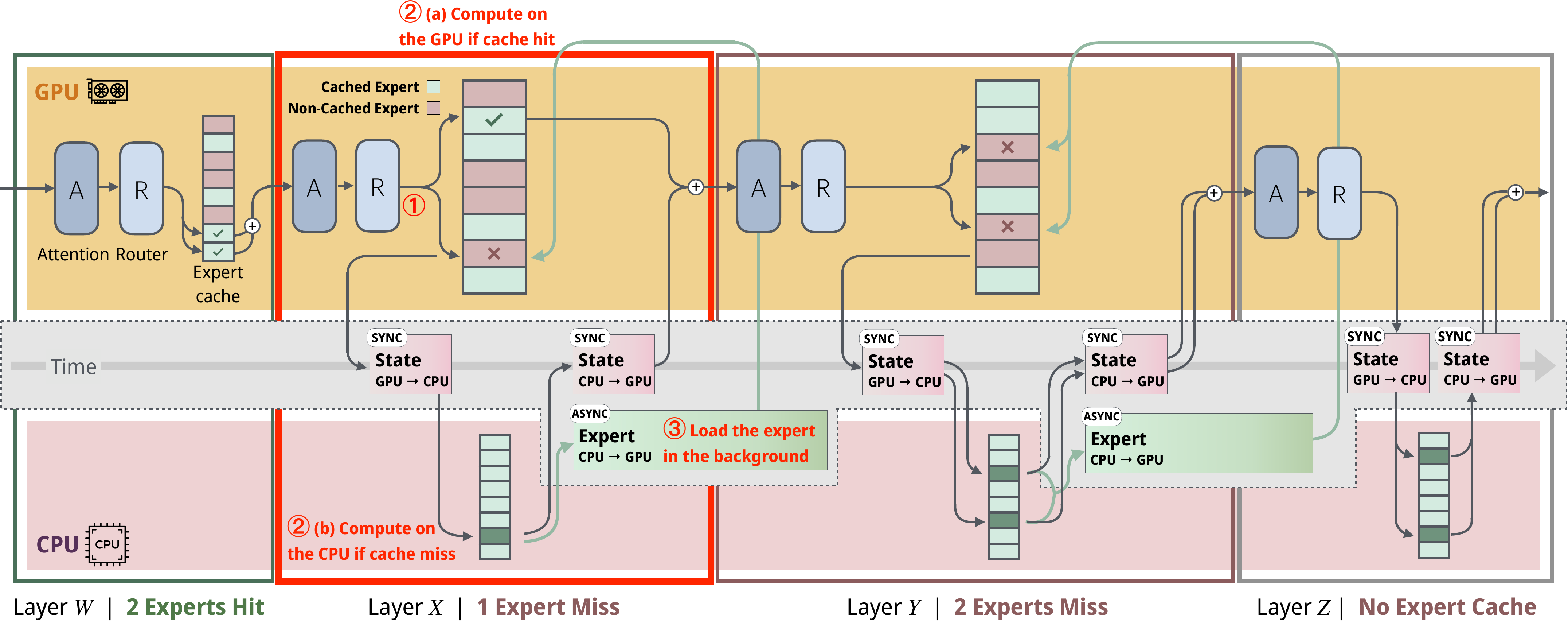}
        \caption{Workflow overview}
        \label{fig:workflow}
    \end{subfigure}
    \begin{subfigure}[t]{0.31\textwidth} 
        \centering
        \raisebox{1.3\baselineskip}{
    \includegraphics[width=\textwidth]{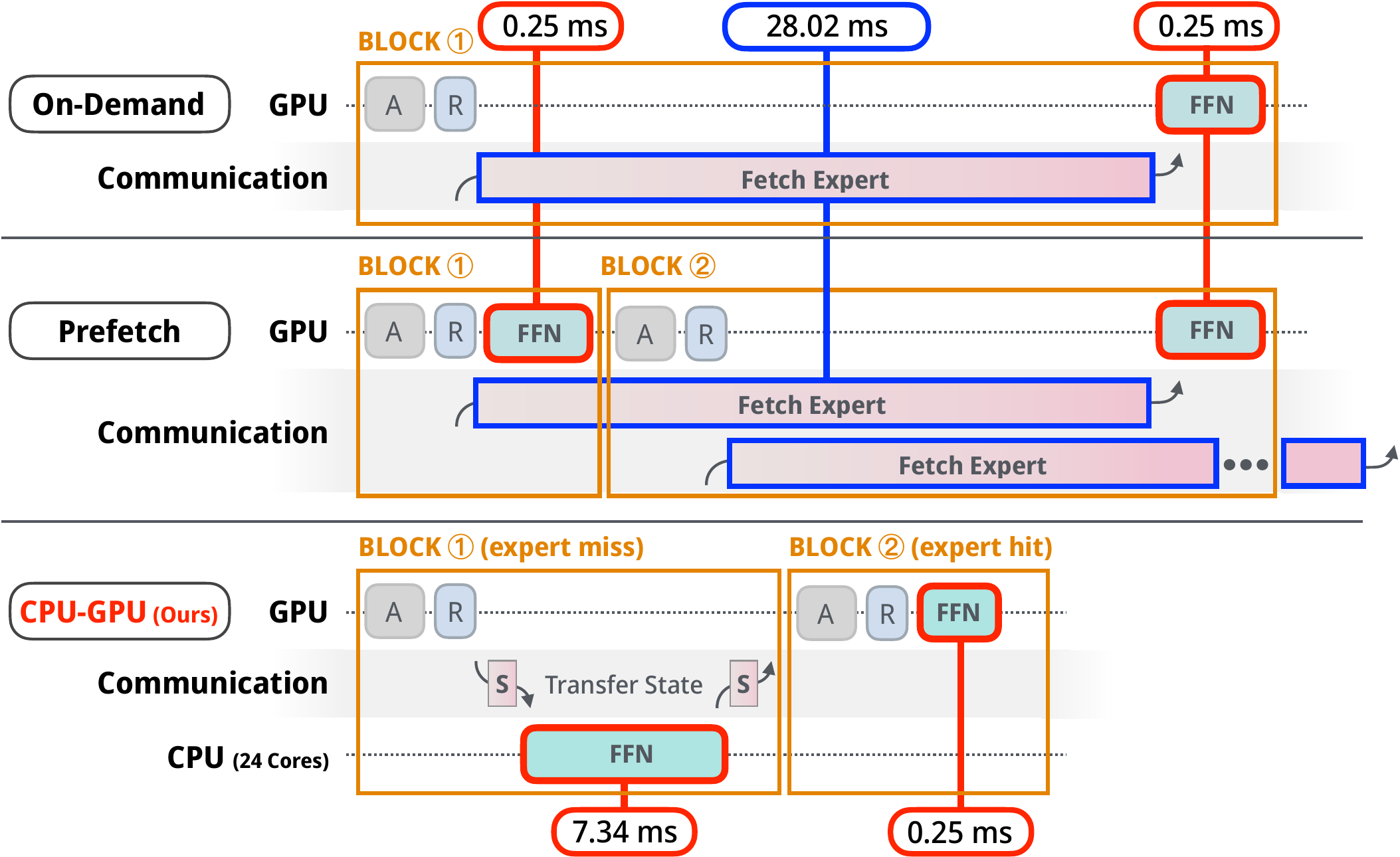}}
        \caption{Timeline comparison}
        \label{fig:timeline}
    \end{subfigure}
    \vspace{-0.5em}
    \caption{
    \cyl{(a) Our workflow, and (b) timing comparison. Workflow includes \protect\circled{\textbf{1}} cache checks, \protect\circled{\textbf{2}} execution based on cache status (\protect\circled{2}(a) on GPU if cache hit, \protect\circled{2}(b) on CPU if cache miss), and \protect\circled{\textbf{3}} asynchronous data transfer to update the GPU cache for future token generation. For layers beyond the cache coverage (e.g., Layer~$Z$), all expert computations are done on the CPU.}
    }
    \vspace{-1.5em}
    \label{fig:method-architecture}
\end{figure*}

\subsection{\cyl{Motivational Observation: Expert Selection Pattern}}
\label{subsec::expert_selection_pattern}


\cyl{Identifying experts that are frequently reused across model layers and consecutive tokens allows us to cache and prefetch portions of these expert weights on the GPU, which enables faster access and improves efficiency. Fig.~\ref{fig:expert-pattern-overview} illustrates two patterns: \textit{Consecutive Layers Pattern} and \textit{Consecutive Tokens Pattern}.}
To explore this potential influence of them, we recorded expert selection results from each layer's router in the Mixtral 8x7B model and analyzed these two patterns.

\textbf{\circled{1} Consecutive Layers Pattern}: \cyl{During each token generation pass, consecutive layers frequently route to identical expert sets. Fig.~\ref{fig:expert-pattern-overview} reveals that approximately $44\%$ of the time, routers direct experts to the same ID as in the previous block. This consistency enables expert selection prediction for weight prefetching. Mixtral-offloading~\cite{2023moefastinf} leverages this pattern through subsequent router networks to predict future expert activations. However, this strategy proves inadequate for offloading scenarios since required expert weights must load within the brief time window of a single layer's forward pass. Similarly, DeepSpeed Inference and Huggingface Accelerate attempt to hide communication latency through sequential-layer preloading but fail to exploit token-level expert reuse across layers, limiting their ability to benefit from repeated expert activations. We identify reuse patterns not only across layers but also across tokens, which we term the \textit{Consecutive Tokens Pattern}. This newly observed behavior enables more effective expert weight prefetching through our caching strategy design in memory-constrained offloading scenarios.}


\textbf{\circled{2} Consecutive Tokens Pattern}: \cyl{Our observations indicate that within the same layer across consecutive token generations, identical experts are frequently selected. Fig.~\ref{fig:expert-pattern-overview} illustrates the frequency of consecutive token generations where identical experts are routed within a given layer. Results demonstrate that at least one expert maintains reuse probability between $40\%$ and $60\%$ across all layers. This pattern suggests performance benefits through GPU-based expert caching, as experts selected for current generation likely serve subsequent tokens. Unlike the \textit{Consecutive Layers Pattern}, which demands caching within a single layer's brief pass, the \textit{Consecutive Tokens Pattern} offers an extended window for expert weight fetching during the current token's forward pass, enabling greater computation-communication overlap.
Further analysis reveals that within each block, the same expert tends to remain active beyond the immediate next token. Among tokens where at least one expert repeats from the previous token, approximately $23\%$ share at least one expert with the previous two tokens, and $18\%$ share with the previous three or more tokens. These patterns indicate high probability that activated experts continue for subsequent tokens.
Rather than relying on offline profiling of static activation distributions, our approach dynamically observes expert reuse patterns and adapts accordingly, enabling applicability across different MoEs. We implement a Least Recently Used (LRU)-style caching strategy guided by the \textit{Consecutive Tokens Pattern}, which reflects the observation that recently selected experts are likely to be reused shortly.}

%% file: sections/3_method.tex
\section{Methodology}
\label{sec:method}



\subsection{Framework Overview}

Our MoE inference framework is developed for optimizing memory usage and computational efficiency by selectively caching expert weights on the GPU. Its essential model components, such as self-attention layers, router networks, and KV caches, reside continuously in the GPU's memory. Given that MoE expert weights constitute the dominant portion of model parameters and memory requirements, the framework allocates the remaining GPU memory to implement an expert cache for storing expert weights. This cache employs an $N$-index, $M$-way set-associative structure spanning layers $0$ through $N-1$ in the MoE model. Each layer maintains capacity for $M$ expert weights to enable rapid access and minimize data transfer needs. Fig.~\ref{fig:method-architecture}~(a) illustrates the workflow, with key steps described as follows:
\textbf{\circled{1} Cache Check:} 
Upon each layer access, the framework checks whether the required experts are available in GPU’s expert cache. 
\textbf{\circled{2} Execution Based on Cache Status:} If an expert resides in the expert cache (i.e., a cache hit), the computation proceeds directly on the GPU so as to minimize communication overhead. For instance, in layer $W$, where all necessary experts are cached, there is no data transfer between CPU and GPU. In contrast, when a cache miss occurs (e.g., in layers $X$ and $Y$), our framework offloads the intermediate state (i.e., the output from the attention layer) to the CPU side for expert computation. 
\textbf{\circled{3} Asynchronous Data Transfer:} For missed experts, the required weights are asynchronously copied to the GPU in the background (i.e., post-fetching) to update the cache for future access.
In the case of cache-missed layers, CPU resources are temporarily leveraged for expert computation while the missing weights are transferred back to the GPU. In addition, for layers beyond the cache coverage (e.g., layer $Z$, where cache use is unavailable due to the insufficient size of available GPU memory, all expert computations are conducted on CPU.

\cyl{Fig.~\ref{fig:method-architecture}~(b) depicts the timeline comparison of our method against on-demand fetching and prefetching methodologies (e.g., Pre-gated MoE~\cite{2024pregatedmoe} and AdapMoE~\cite{2024adapmoe}). Prefetching method efficiency remains limited since computation time is significantly less than communication cost. Although visualization suggests overlapping expert transfers in Pre-gated MoE, their actual implementation of data transfers remains serialized\footnote{They created a single CUDA stream for the fetcher (i.e., the structure that fetches required data from CPU to GPU) in the released source code, so all communications on the same CUDA stream are performed sequentially.}. In contrast, our collaborative framework reduces data transfer by offloading expert computations to CPU, enabling accelerated token generation. Meanwhile, the expert cache and asynchronous transfers further ensure efficient GPU utilization as data transfers can be overlapped with computations.}


\subsection{Memory Management and Cache Design}
\label{sec:method:mem}

\cyl{Our framework divides GPU memory into two sections: an expert cache and a section for other essential model components, including weights for self-attention, normalization, router networks, and the KV cache. These components typically require a small amount of memory (e.g., Mixtral 8x7B requires approximately 5 GB for them), with remaining GPU memory allocated to the expert cache. Expert cache configuration depends on available GPU memory and individual expert size. The number of expert slots ($S$) is determined by dividing the available GPU memory by the size of an individual expert as }$
    \label{eq:gpu_slots}
    S = \left\lfloor\frac{\text{available GPU memory}}{\text{per-expert size}}\right\rfloor
$.
The number of expert slots represents the total cache capacity in terms of expert count. Through specification of the number of ways per layer $M$, the number of indexes ($N$) in the cache is calculated as: $
    N=\left\lfloor\frac{S}{M}\right\rfloor $.
For example, on an NVIDIA GeForce RTX 4090 with 24 GB of memory, around 56 expert slots can be accommodated for the Mixtral 8x7B model, which consists of 256 experts in total (i.e., eight experts per layer across 32 layers). Configuring the GPU memory as a 4-way set-associative cache results in 14 indexes (i.e., $\frac{56}{4} = 14$). Such a configuration enables expert coverage for layers 0 to 13.

\cyl{Our expert cache employs an LRU eviction policy. Due to the \textit{Consecutive Tokens Pattern} property discussed in Section~\ref{subsec::expert_selection_pattern}, this LRU policy ensures that frequently accessed experts remain in the cache.
Our experiments also evaluated the LRU policy against other strategies, including First In, First Out (FIFO) and random selection, all in Section~\ref{sec:exp:cache}.}


\subsection{Efficient Transfer \& Implementation Details}

To enhance data transfer performance, our framework leverages NVIDIA's dual-copy engine capability, which enables concurrent data transfers. This parallel transfer capability is implemented through two independent CUDA streams. Without these dual streams, data transfers would be serialized, creating a performance bottleneck, as transfers of intermediate states (i.e., attention outputs) from GPU to CPU could need to wait for ongoing expert weight transfers to complete.

Our CPU-GPU collaborative framework builds upon the official MistralAI repository~\cite{mistralgithub}, which implements the Mixtral 8x7B models. We furthermore adapted this codebase to support Hugging Face-formatted models (e.g., Phi3.5-MoE).

%% file: sections/4_exp.tex
\section{\cyl{Experimental Results and Insights Discussion}}
\label{sec:exp}


\subsection{Experimental Setup}
\label{sec:exp:setup}
\begin{table}[t]
    \centering
    \caption{Configuration comparison of two MoE-based LLM models: Mixtral 8x7B and Phi3.5-MoE.}
    \label{table:model-comparison}
    \resizebox{0.42\textwidth}{!}{
        \begin{tabular}{lcc}
        \toprule
        Model                       & Mixtral 8x7B~\cite{jiang2024mixtralexperts} & Phi3.5-MoE~\cite{2024phi3.5} \\
        \midrule
        \# of parameters            & 46.7B        & 41.9B  \\
        Model size                  & 88 GB        & 79 GB   \\
        \# of layers                & 32           & 32        \\
        \# of experts per layer     & 8            & 16         \\
        Select top-K experts        & 2            & 2          \\
        Size per expert             & 340 MB       & 152 MB \\
        \bottomrule
        \end{tabular}
    }
\end{table}

\begin{figure}[t]
    \centering
    \begin{subfigure}[b]{\linewidth}
        \centering
        \includegraphics[width=0.94\linewidth]{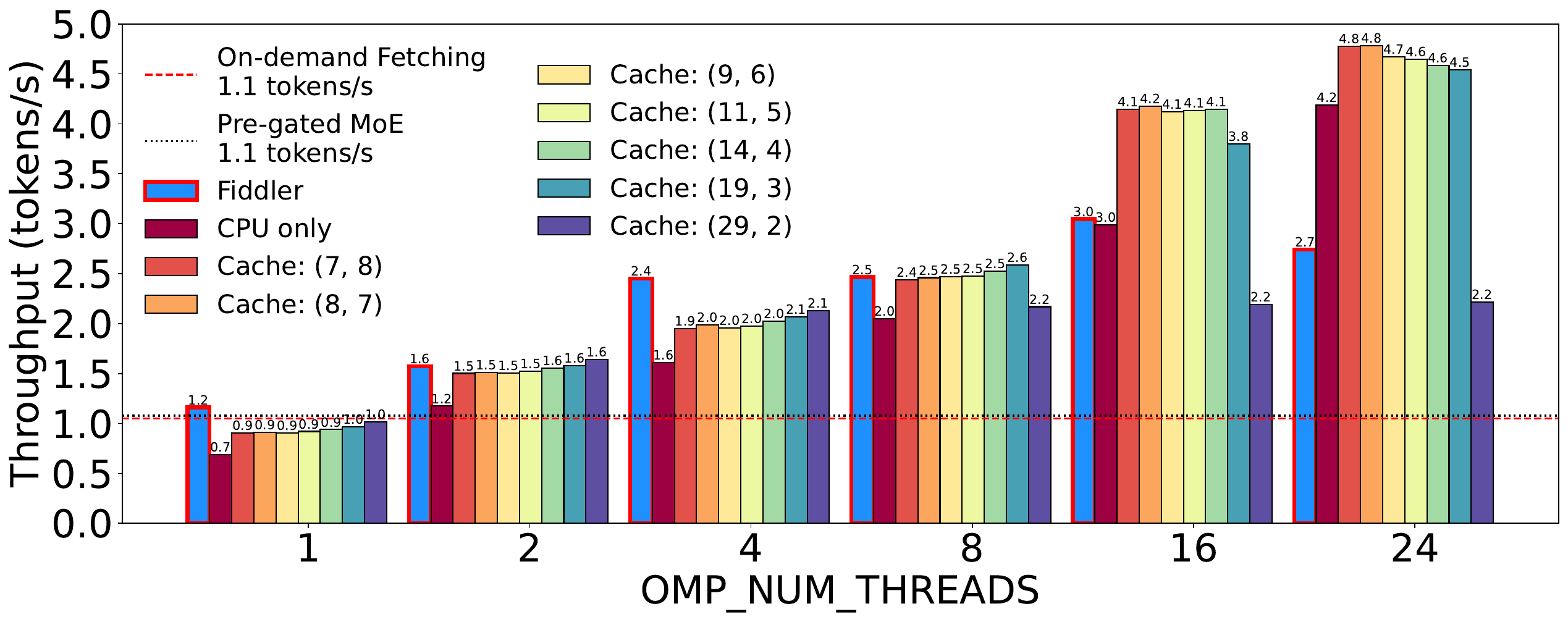}
        \vspace{-0.5em}
        \caption{Mixtral 8x7B}
        \label{fig:mixtral-overall}
    \end{subfigure}
    \vskip 0.5em
    \begin{subfigure}[b]{\linewidth}
        \centering
        \includegraphics[width=0.94\linewidth]{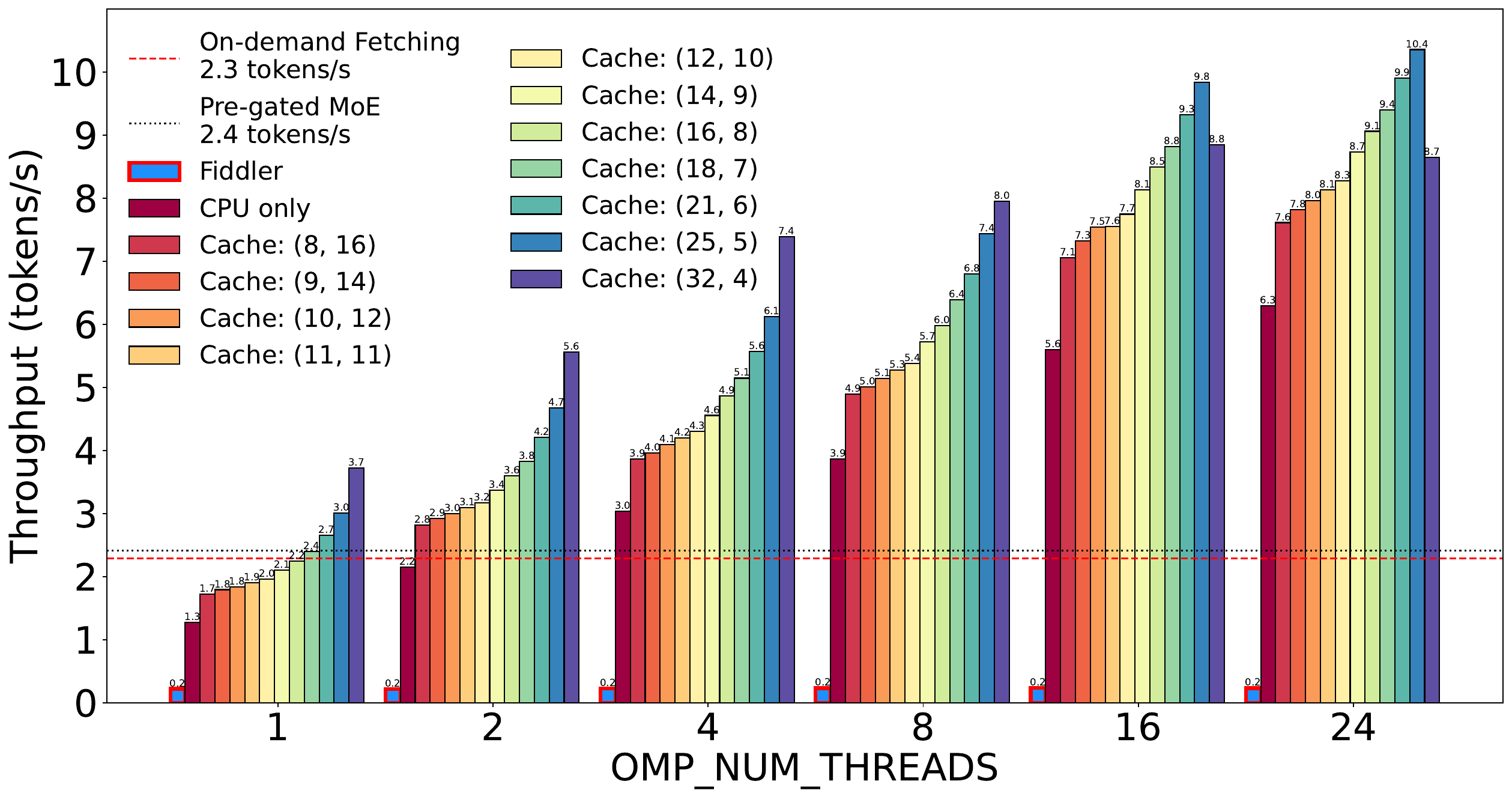}
        \vspace{-0.5em}
        \caption{Phi3.5-MoE}
        \vspace{-0.5em}
        \label{fig:phi3.5-overall}
    \end{subfigure}
    \caption{Overall performance comparison of (a) Mixtral 8x7B and (b) Phi3.5-MoE models under different number of CPU cores (\texttt{OMP\_NUM\_THREADS}) and cache configurations. The cache legend indicates (\# of indexes, \# of ways).}
    \label{fig:overall-performance}
    \vspace{-0.5em}
\end{figure}


We compare our proposed framework against on-demand expert fetching~\cite{deepspeedinference,huggingfaceTransformers} expert prefetching framework, Pre-gated MoE~\cite{2024pregatedmoe}, and a CPU-GPU collaboration framework, Fiddler~\cite{fiddler2025}, on two state-of-the-art MoE-based LLMs: Mixtral 8x7B~\cite{jiang2024mixtralexperts} and Phi3.5-MoE~\cite{2024phi3.5}. Since Pre-gated MoE implementation does not support these newer models, we estimate its performance by assuming perfect computation-communication overlap. Table~\ref{table:model-comparison} compares the Mixtral 8x7B and Phi3.5-MoE models. Both models select two experts per layer for inference; however, Phi3.5-MoE employs smaller expert sizes that result in reduced activated parameters. The expert cache capacity in slot count is determined by available GPU memory and individual expert size, according to the details in Section~\ref{sec:method:mem}.
Our experiments are conducted on a system equipped with an AMD Ryzen Threadripper 7960X 24-core CPU and NVIDIA RTX 4090 24GB GPU. The maximum CPU frequency varies between 4.8GHz and 5.3GHz based on active core count, where higher core usage reduces frequency. The GPU connects to the CPU via PCIe Gen 4.0 ×16 interconnect. Our power analysis monitors CPU package power through Running Average Power Limit (RAPL) and GPU power through the \texttt{nvidia-smi} command.


\begin{table}[t]
    \centering
    \caption{Expert computation and communication time (in milliseconds) for the  Mixtral 8x7B and Phi3.5-MoE models.}
    \label{table:ffn-time}
    \resizebox{0.48\textwidth}{!}{
        \begin{tabular}{l|l|c|cccccc}
        \toprule
        \multirow{2}{*}{Model}      & Device             & GPU      & \multicolumn{6}{c}{CPU (w/ different no. of threads)}                     \\
                                    & CPU threads  & -        & 1 & 2 & 4 & 8 & 16 & 24 \\
        \midrule
        \multirow{2}{*}{Mixtral 8x7B} & Expert comp. time & 0.25   & 44.12 & 25.53 & 18.34 & 15.76 & 10.96 & 7.34 \\
                                    & Comm. time & 28.02    & \multicolumn{6}{c}{0.11}                    \\
        \midrule
        \multirow{2}{*}{Phi3.5-MoE}   & Expert comp. time & 0.11   & 22.73 & 12.80 & 8.58 & 6.39 & 3.92 & 3.36     \\
                                    & Comm. time & 12.26    & \multicolumn{6}{c}{0.11}                    \\
        \bottomrule
        \end{tabular}
    }
    \vspace{-1em}
\end{table}

\subsection{Overall Performance Analysis}
\label{sec:exp:overall}

Fig.~\ref{fig:overall-performance} presents a comprehensive performance comparison of our proposed CPU-GPU collaborative inference framework with on-demand expert fetching~\cite{deepspeedinference,huggingfaceTransformers}, the prefetching strategy (Pre-gated MoE)~\cite{2024pregatedmoe} and Fiddler~\cite{fiddler2025} under varying numbers of CPU cores. The \textit{CPU only} configuration refers to the scenario where no GPU expert cache is employed, and all expert computations occur on the CPU. With two to four CPU cores, the CPU-only version surpasses the prefetching works, as the reduced CPU computation time becomes shorter than the PCIe transfer time for expert weights. This configuration, where all computations are performed on the CPU, establishes the performance lower bound for our method, corresponding to a scenario with a 100\% cache miss rate. Furthermore, with the employment of the GPU expert cache, our framework achieves approximately $15\%\sim 35\%$ and $50\%\sim250\%$ performance improvements over the \textit{CPU only} setup for Mixtral 8x7B and Phi3.5-MoE, respectively.

\cyl{Our framework significantly outperforms existing methodologies, achieving up to 4.4$\times$ speedup for Mixtral 8x7B and 4.3× for Phi3.5-MoE with full 24-core CPU utilization, while surpassing Fiddler by approximately 1.6$\times$. Note that Fiddler performs poorly on Phi3.5-MoE due to its reliance on offline profiling of expert popularity, which limits applicability to other MoE models. Furthermore, the increased expert count in Phi3.5-MoE introduces significant overhead for Fiddler's CPU-GPU execution decisions due to exponential complexity.}

Our collaborative CPU-GPU framework significantly boosts inference throughput by efficiently coordinating resources. Even with minimal CPU cores, like a single one, our method outperforms Pre-gated MoE by 54\% on the Phi3.5-MoE model. With just two CPU cores, it consistently surpasses previous methods on both tested models, showcasing its efficiency across diverse resource configurations. This flexibility means our framework is not limited to high-core systems and can adapt to various hardware setups.
Furthermore, our approach does not require any model modifications and dataset profiling, ensuring compatibility with new MoE models. While Pre-gated MoE  necessitates router network changes, complicating reproducibility and accuracy evaluations, Fiddler requires profiling on dataset to select which experts are placed on the GPU. Additionally, Fiddler's algorithm for determining an expert model's execution unit (e.g., one the CPU or the GPU) can serve as a bottleneck for models with a large number of experts due to its exponential complexity. We offer a straightforward integration and robust solution for enhancing inference performance of MoE model variants.

\subsection{\cyl{Performance vs CPU Count and Cache Configurations}
}
\label{sec:exp:perfimpact}

Table~\ref{table:ffn-time} provides a detailed breakdown of the expert's computation time and communication time for the Mixtral 8x7B and Phi3.5-MoE models across different CPU core counts. These evaluation results, along with Fig.~\ref{fig:overall-performance}, reveal two primary performance trends as follows:

\textbf{With Fewer CPU Cores (1$\sim$4):} In low-core configurations, expert computation time per core is significantly higher than communication time. For example, with a single CPU core, the computation time for an expert layer in Mixtral 8x7B  reaches 44.12 ms, whereas the computation time on the GPU is only 0.25 ms. This imbalance highlights expert's computation as the primary bottleneck. As a result, cache configurations with a higher number of indexes and fewer ways are optimal in this scenario. By maximizing expert reuse on the GPU and covering more layers, this setup achieves greater speedup to effectively offset the high computation cost on the CPU. Nevertheless, the trade-off of using fewer cache ways results in an increased cache miss rate, which leads to more frequent CPU-GPU transfers. This trade-off is favorable in low-core scenarios since the computation dominates the total latency to make the added communication overhead less impactful. However, this configuration becomes less advantageous as the number of CPU cores increases.

\textbf{With Increased CPU Cores (8$\sim$24):} As the number of CPU cores increases, expert's computation time decreases significantly due to parallelism. For instance, the computing time for Mixtral 8x7B drops from 44.1 ms on one core to 7.3 ms on 24 cores. Under these high-core conditions, communication overhead in low-way cache configurations becomes the primary bottleneck, as the reduced computation time cannot fully mask data transfer latency. In the Mixtral 8x7B model, for example, a low-way configuration like (29,2) caps performance at around 2.2 tokens/second to indicate that data transfer rather than computation is limiting further gains.
Therefore, cache configurations with more ways and fewer indexes are more efficient. This structure improves cache hit rates and reduces the frequency of CPU-GPU transfers,  minimizing data transfer overhead. These trends highlight the importance of cache adjustments to balance CPU utilization and transfer efficiency as cores scale up.

\subsection{Cache Eviction Policy Analysis}
\label{sec:exp:cache}

\begin{figure}[t]
    \centering
    \begin{subfigure}[b]{0.45\textwidth}
        \centering
        \includegraphics[width=\textwidth]{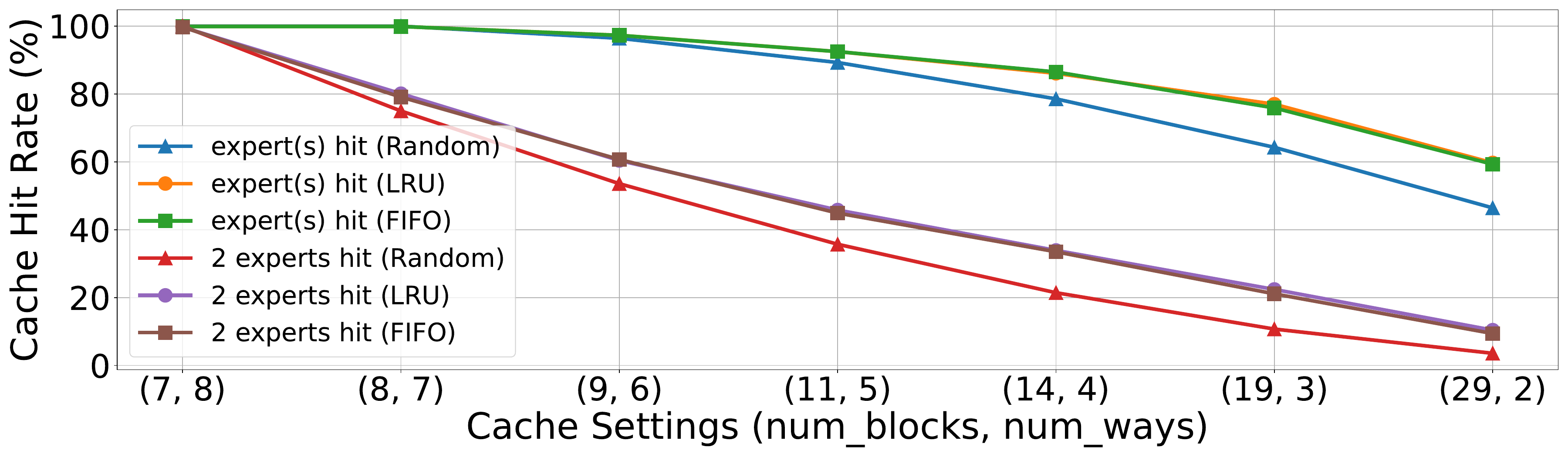}
        \vspace{-1.8em}
        \caption{Mixtral 8x7B}
        \label{fig:mixtral-cache}
    \end{subfigure}
    \begin{subfigure}[b]{0.45\textwidth}
        \centering
        \includegraphics[width=\textwidth]{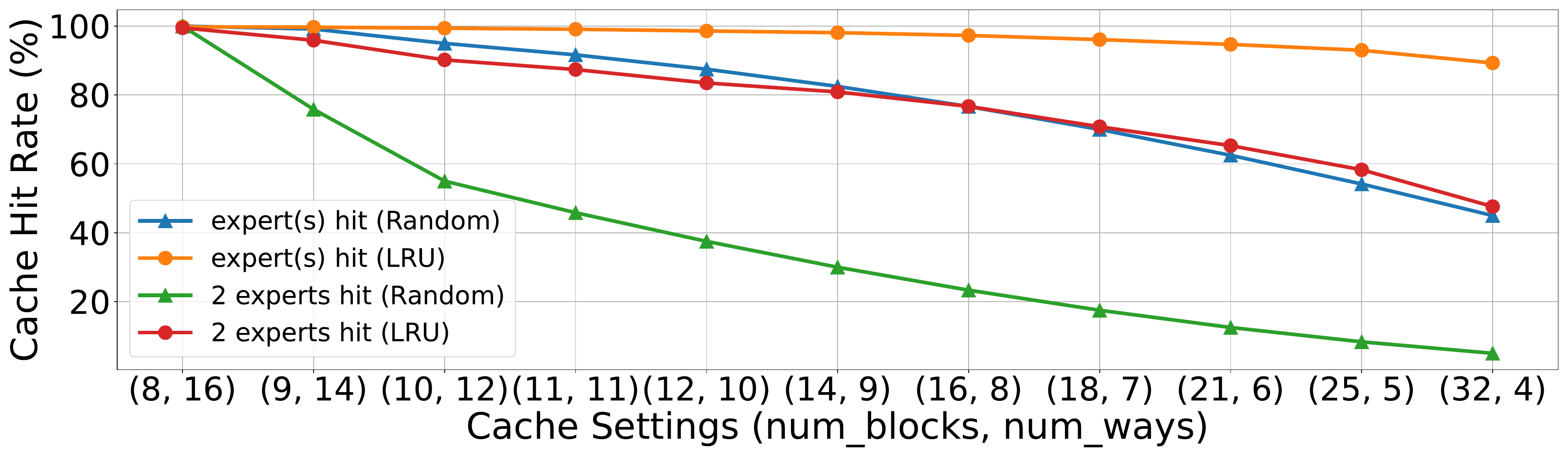}
        \vspace{-1.8em}
        \caption{Phi3.5-MoE}
        \vspace{-0.5em}
        \label{fig:phi3.5-cache}
    \end{subfigure}
    \caption{Expert cache hit rate comparisons for two MoE-based LLM models: (a) Mixtral 8x7B and (b) Phi3.5-MoE under different expert cache configurations and eviction policies.}
    \label{fig:overall-cache}
\end{figure}

Fig.~\ref{fig:overall-cache} illustrates the cache hit rate under various cache configurations and eviction policies for Mixtral 8x7B and Phi3.5-MoE. Two types of hit rates are presented: ``expert(s) hit'', indicating at least one expert is present in the cache, and ``2 experts hit'', indicating both experts selected by the router network are present. The random policy involves randomly selecting a set of expert networks to be stored in the cache statically. Since these experts remain fixed in the cache, the hit rates can be calculated using the following equations: 
$
P(\text{At least 1 expert hit}) = 1 - \left(\frac{n - M}{n}\right)\times\left(\frac{n - M - 1}{n - 1}\right), P(\text{2 experts hit}) = \left(\frac{M}{n}\right)\times\left(\frac{M - 1}{n - 1}\right),
$
where $M$ represents the number of cache ways for each index, and $n$ denotes the number of experts per layer. These are labeled as ``expert(s) hit~(Random)” and ``2 experts hit~(Random)” in Fig.~\ref{fig:overall-cache}, respectively. In Fig.~\ref{fig:mixtral-cache}, the experimental results indicate that both LRU and FIFO policies perform similarly, with LRU slightly outperforming FIFO. As a result, Fig.~\ref{fig:overall-cache}~(b) only incorporates a comparison between LRU and the random policy to avoid potential misleading interpretations.

The results in Fig.~\ref{fig:overall-cache} demonstrate that the LRU replacement policy in the expert cache outperforms the random policy. For Mixtral 8x7B, LRU improves the probability by around $5\sim15\%$, while for Phi3.5-MoE, LRU produces significantly better results than random selection. These improvements stem from two main factors: (1) Although Phi3.5-MoE contains twice the number of experts, this also increases the number of ways, which enables the expert cache to accommodate more unique experts. (2) The expert selection tends to remain stable between consecutive forward passes (e.g., generating two tokens). For example, if experts 0 and 1 were chosen in the previous pass, it is probable that one of them would be reused in the next pass, along with a new expert. The increased number of ways also makes it more likely that this new expert would be in the cache.  This analysis of our cache eviction policy highlights the effectiveness of our expert-fetching mechanism, even without immediate fetching for each forward pass, as seen in methods such as~\cite{2024pregatedmoe} and other prefetching-based techniques~\cite{2023moefastinf,2024adapmoe,2024expertflow}. By leveraging the CPU for expert computations alongside an proper cache eviction policy, our framework effectively balances expert reuse with cache management to achieve improved overall inference performance.

\subsection{Power and Energy Analysis}
\label{sec:exp:power}

\begin{table}[t]
    \centering
    \caption{Power (W) comparison of our method with different numbers of CPUs against Pre-gated MoE~\cite{2024pregatedmoe}.}
    \label{table:power}
    \resizebox{0.48\textwidth}{!}{
        \begin{tabular}{l|c|cccccc|c}
        \toprule
        \multirow{2}{*}{Model}   &\multirow{2}{*}{Device}& \multicolumn{6}{c|}{(\textbf{Ours}) OMP\_NUM\_THREADS}        & \multirow{2}{*}{Prefetching~\cite{2024pregatedmoe}}\\
                                    &     & 1    & 2     & 4     & 8     & 16    & 24    &               \\
        \midrule
        \multirow{2}{*}{Mixtral 8x7B} & CPU & 86.1 & 91.7  & 100.3 & 111.0 & 133.4 & 147.5 & 92.1          \\
                                    & GPU & 91.6 & 92.8  & 101.0 & 103.4 & 99.6  & 97.9  & 96.3          \\
        \midrule
        \multirow{2}{*}{Phi3.5-MoE}   & CPU & 84.4 & 88.4  & 92.0  & 98.4  & 110.1 & 118.3 & 88.2          \\
                                    & GPU & 97.4 & 100.7 & 105.9 & 109.2 & 106.0 & 109.2 & 100.7        \\
        \bottomrule
        \end{tabular}
    }
    \vspace{-0.5em}
\end{table}

\begin{table}[t]
    \centering
    \caption{Joule-per-token comparison of our method with different CPU cores against Pre-gated MoE~\cite{2024pregatedmoe}.}
    \vspace{-0.5em}
    \label{table:energy}
    \resizebox{0.48\textwidth}{!}{
    \begin{tabular}{l|cccccc|c}
        \toprule
        \multirow{2}{*}{Model} & \multicolumn{6}{c|}{(\textbf{Ours}) OMP\_NUM\_THREADS}                & \multirow{2}{*}{Prefetching~\cite{2024pregatedmoe}} \\
                                        & 1       & 2      & 4      & 8      & 16     & 24     &\\
        \midrule
        Mixtral 8x7B                       & 177.7 & 115.3 & 95.8 & 82.5 & 55.5 & 51.1 & 171.3 \\
        Phi3.5-MoE                         & 49.1  & 33.8 & 26.7 & 25.9 & 22.1 & 21.9 & 78.7 \\
        \bottomrule
        \end{tabular}
    }
    \vspace{-1em}
\end{table}


Table~\ref{table:power} presents the power consumption of our collaborative approach across different CPU core counts, compared to the prefetching method~\cite{2024pregatedmoe}. For each CPU core configuration, we select the optimal cache setup based on findings in Section~\ref{sec:exp:overall}. In our method, CPU power consumption is generally higher than in previous methods due to increased CPU core utilization. On average, each core consumes approximately $2\sim4$ Watts, as measured by RAPL.
Computing expert networks directly on the CPU and leveraging expert caching reduces the frequency of CPU-GPU communication and higher power consumption, indicating that our framework effectively maximizes available resources.

Regarding overall energy consumption, our reduced generation time enables a substantial reduction. Although increased CPU utilization leads to higher power draw, it is significantly more energy-efficient than prefetching methods. Table~\ref{table:energy} illustrates the energy required for generating a token, showing that
our method demonstrates significant energy savings across all CPU core configurations. For instance, with 24 CPU cores, our method consumes only $29.9\%$ of the energy required by the prefetching method for the Mixtral 8x7B model and $27.8\%$ for the Phi3.5-MoE model.
The framework's ability to adapt to different CPU core counts ensures that it remains energy-efficient across various hardware configurations.

%% file: sections/5_conclusion.tex
\section{Conclusion}
\label{sec:conclusion}

We introduced a novel CPU-GPU collaborative inference framework for MoE-based language models on memory-limited GPU systems. Our approach efficiently leverages CPU multi-core parallelism for expert computations and utilizes GPU memory as an expert cache with an asynchronous fetching policy, outperforming prefetching methods by overlapping computations and communication. Validated on Mixtral 8x7B and Phi-3.5-MoE, our framework achieved up to 4.8 and 10.4 tokens/second, respectively, demonstrating significant speedups (up to 4.4$\times$) and improved power efficiency.

%% file: sections/6_acknowledgements.tex
\section{Acknowledgements}
The authors gratefully acknowledge the support from the National Science and Technology Council (NSTC) in Taiwan under grant numbers NSTC 114-2221-E-002-069-MY3, NSTC 113-2221-E-002-212-MY3, NSTC 114-2218-E-A49-026, and NSTC 114-2640-E-002-006. The authors would also like to express their appreciation for the support from Google Inc. Furthermore, the authors extend their gratitude to the National Center for High-Performance Computing (NCHC) for providing the necessary computational and storage resources.